\documentclass[aps,floatfix]{revtex4}
\usepackage{epsfig}
\usepackage{color}
\usepackage{amssymb}
\usepackage{amsmath}
\usepackage[ngerman,english]{babel}
\usepackage{multirow}
\usepackage{graphicx}
\usepackage{subfigure}

\begin{document}

\title{Light element ($Z=1,2$) production from spontaneous ternary fission of $^{252}$Cf}
\author{G. R\"{o}pke}
\email{gerd.roepke@uni-rostock.de}
\affiliation{Institut f\"{u}r Physik, Universit\"{a}t Rostock, D-18051 Rostock, Germany}
\author{J. B. Natowitz}
\email{natowitz@comp.tamu.edu}
\affiliation{Cyclotron Institute, Texas A\&M University, College Station, Texas 77843, USA}
\author{H. Pais}
\email{hpais@uc.pt}
\affiliation{CFisUC,  Department of Physics, University of Coimbra, 3004-516 Coimbra, Portugal}

\date{\today}

\begin{abstract}
The yields of light elements ($Z=1,2$) obtained from spontaneous ternary fission of $^{252}$Cf are treated within a nonequilibrium approach, and the contribution of unstable nuclei and excited bound states is taken into account. These light cluster yields may be used to probe dense matter, and to infer in-medium corrections. Continuum correlations are calculated from scattering phase shifts using the Beth-Uhlenbeck formula, and the effect of medium modification is estimated. The relevant distribution is reconstructed from the measured yields of isotopes. This describes the state of the nucleon system at scission and cluster formation, using only three Lagrange parameters which are the nonequilibrium counterparts of the temperature and chemical potentials, as defined in thermodynamic equilibrium. We concluded that a simple nuclear statistical equilibrium model neglecting continuum correlations and medium effects is not able to describe the measured distribution of H and He isotopes. Moreover, the freeze-out concept may serve as an important ingredient to the nonequilibrium approach using the relevant statistical operator concept.
\end{abstract}


\maketitle

\section{Introduction}

Thermal neutron induced and spontaneous ternary fission is a process in which the emission of two medium-mass fragments is accompanied by  equatorially emitted  light particles and clusters formed in the neck region at the time of scission, see \cite{[1],[2],[3]} and references therein. Data for cluster yields obtained from ternary fission experiments with thermal neutrons are shown, e.g., in \cite{Koester,Koestera,KoesterPhD}.
In particular, data for the ternary fission yields of $^{241}$Pu($n_{\rm th}$,f) are presented. An interpretation of the Koester {\it et al.} data \cite{Koester}  was given in Ref. \cite{Sarah}, where a suppression of the yields of the larger clusters is found, due to cluster formation kinetics.
Also, ternary fission has been observed
 from other actinides such as $^{239}$Pu, $^{233}$U, $^{235}$U, and $^{245}$Cm. For more recent work on ternary fission see \cite{Zagrebaev,Artemenkov,Holmvall,Chuvilskaya,Wollersheim}. 

Different approaches have been employed to interpret these data, see  \cite{Koestera}, and often a Boltzmann distribution has been used. 
An interpolation formula has been presented in \cite{Valskii2} which describes the general behavior of the measured yields but cannot explain the details of the observed distributions.
More fundamentally, the use of a Boltzmann distribution as known from thermodynamic equilibrium with parameters  temperature and chemical potentials remains unfounded.
However, the signatures of binding energies and degeneracy of the isotopes according to the  Boltzmann distribution are clearly seen in the observed yields.

In this paper we report on investigations of the yields of equatorial  emission of $Z=1,2$ isotopes during ternary spontaneous fission of $^{252}$Cf. We are interested in a better understanding  of cluster formation and the fate of correlations in low density, low-temperature, expanding nuclear matter.  
This nonequilibrium evolution can be described using the method of the nonequilibrium statistical operator (NSO) \cite{NSO}.
It is based on information-theoretical concepts which is also the basis of equilibrium statistical physics. 
We include different processes which describe the dynamical evolution of the system.
In particular, we include the decay from other unstable nuclei (feeding), 
the inclusion of excited states (including continuum correlations), and medium effects. 
For this, a quantum statistical approach is used.

Our main goal is to shed some more light into the properties of dense nuclear matter, namely, to understand how to improve the simple model of nuclear statistical equilibrium (NSE)
by including continuum correlations and  in-medium effects, and how
 well the nucleon density at the time of onset of cluster formation can be determined. We discuss the apparent suppression of yield for some exotic isotopes and the relation to thermodynamic quantities.

The paper is organized as follows: in Section  \ref{sec:exp}, we collect available data on ternary spontaneous fission of $^{252}$Cf  concerning the production of the lightest elements H and He. In Sec.\ref{sec:int}, the theoretical formalism used in the work, based on information theory, will be described. Section \ref{sec:med} analyses different numerical calculations that go beyond the ideal gas description considering continuum correlations and in-medium effects. The use of a local density approximation is discussed in Section \ref{sec:LDA}, and finally in Section  \ref{sec:therm}, some final remarks, namely, the relation to thermodynamics, are drawn.

\section{Experimental data}
\label{sec:exp}

A number of experimental investigations of  the spontaneous fission of $^{252}$Cf have been performed
 by different groups, see \cite{KoesterPhD} and references given there.  It has been demonstrated  that $^{252}$Cf(sf) emits $3.2(1) \cdot 10^{-3}$ $\alpha$-particles per fission \cite{Rai68,Wil85}.  Usually, the yields $Y_{A,Z}$ of other light elements produced from ternary fission are normalized to the final yield of $^4$He ($\alpha$), 
 which is fixed to $Y_{^4{\rm He}^f}=10000$. More precisely, we consider here the ratios of yields relative to the final yield of $\alpha$-particles.
 
 In Ref. \cite{KoesterPhDtab}, the author performed a compilation of results for ternary fission data, taken from the literature. 
There,  consistent experimental results are presented for the $^1$H ($p$) and $^2$H ($d$)  yields, as well as for the $^8$He ones.
However, the values for the yield of  $^3$H ($t$), and also of  $^6$He, are quite different. These data have been used also 
in a more recent publication \cite{Valskii2}, where the  yields for the H isotopes are from Ref.  \cite{Whe67}, and  for the He isotopes, the yields are according Ref. \cite{Dlo92}.

\begin{table}
\begin{center}
\hspace{0.5cm}
 \begin{tabular}{|c|c|c|c|c|c|c|c|c|c|c|c|c|c|c|}
\hline
isotope& $A$  &  $Z$ & $Y^{\rm exp}_{A,Z}$& $Y^{\rm interp}_{A,Z}$&  $\frac{B_{A,Z}}{A} $& $ g_{A,Z} $ & $Y^{\rm final}_{A,Z}$ & $E^{\rm thresh}_{A,Z}$  &$R^\gamma_{AZ}(1.3)$& $Y^{\rm rel,\gamma}_{A,Z}$ 
& $R_{A,Z}^{\rm vir}(1.3)$ & $Y^{\rm rel, vir}_{A,Z}$ & $R_{A,Z}^{\rm eff}(1.3)$ & $Y^{\rm rel, eff}_{A,Z}$\\
\hline
 $\lambda_T$ 	& -&-&-&(1.25)	&-&-		 & 0.806219 &-&-& 1.28018 &- & 1.25052 & - & 1.30556  \\
$\lambda_n$ 		& -&-&-&(-2.8)	&-&-		 & -2.19217 &-&-& -3.18741 &- & -3.1107 & - &-3.00665 \\
$\lambda_p$ 		& -&-&-&(-15.8)&-&-		 & -17.7023 &-&-& -16.5882 &- & -16.7538 & - &-16.6463 \\
\hline
$^1$n		&1& 0	&-			&0.65e6 	&0		& 2 &2.572e8             &-               &-               & 1.084e6        &-              &1.647e6  &-              & 954222\\
$^1$H		&1& 1 	& 160(20)		& 19.4	&0		& 2 & 1.136 	        	&-               &-               & 30.8231         &-              & 30.096 	&-              & 27.625 \\
$^2$H 		&2& 1 	& 63(3) 		& 37.5	& 1.112 	& 3 & 5.012		&2.224	&1		& 61.6143		& 0.98	& 61.579 	& 0.98	& 63.004 \\
$^3$H$^f$	&3& 1 	& 950(90)		& 591	& 2.827 	& 2 & 950			&-		&-		& 951.351		& -		& 951.34 	& -		& 950.06 \\
$^3$H		&3& 1 	& -			&-		& 2.827 	& 2 & {\it (950)}		&6.257	&1		& 829.985		& 0.99	& 943.12 	& 0.99	& 938.91 \\
$^4$H		&4& 1 	& -			& -		& 1.720	& 5 & {\it  (52.95 )}	&-1.6	&1.473	& 121.366		&0.0606	& 8.2191  	& 0.0606	& 11.154 \\
$^3$He		&3& 2 	& -			&  0.0095	& 2.573	& 2 &1.624e-6		&-		&1		& 0.012984 	& 0.988	& 0.00933	& 0.988	& 0.01511 \\
$^4$He$^f$	&4& 2 	& 10000		& 10028	& 7.073 	& 1 & 10000  		&-		&- 		& 10000  		& - 		&10000  	& - 		&10000\\
$^4$He		&4& 2 	& 8264(341)	& -		& 7.073 	& 1 &  {\it (10000)}  	&20.577	&1 		& 7920.86  	& 1 		& 8454.0 	& 1 		& 8256.7 \\
$^5$He		&5& 2 	& 1736(274)	& 2600	& 5.512	& 4 &  {\it (1487)} 	&-0.735	&1		& 2072.8  		&0.7044 	&1540.9 	& 0.6596 	&1733.9 \\
$^6$He$^f$	&6& 2 	& 270(30)		& 369	& 4.878 	& 1 &  270  		& -		&-		& 280.724  	& - 		& 280.6	 & - 		& 270.02  \\
$^6$He		&6& 2 	& 223(26)		& -		& 4.878 	& 1 &   {\it (270)}  	& 0.975	&1		& 215.488  	& 0.9453 	& 222.4  	&0.6961 	& 223.16 \\
$^7$He		&7& 2 	& 47(9)		& 128	& 4.123	& 4 &  {\it (53.76) } 	&-0.410 	&1		& 65.2355  	& 0.821 	& 58.16 	&0.3988 	& 46.863 \\
$^8$He$^f$	&8& 2 	& 25(5)		& 30.6	& 3.925 	& 1 & 25 			& -		&-		& 12.8212  	&- 		& 13.58 	&- 		& 25.110 \\
$^8$He		&8& 2 	& 25(5)		& - 		& 3.925 	& 1 &  {\it (25)}		& 2.125	&1		& 11.9324 	&0.9783 	& 13.32 	&0.9783 	& 24.513 \\
$^9$He		&9& 2 	& -			& - 		& 3.349	& 2 &  {\it (0.8335)}  	& -1.25	&1		& 0.88875   	& 0.2604 	&0.258 	& 0.2604  &0.59706 \\
$^8$Be		&8& 4	& 10(6)		& 15.2	& 7.062 	& 1 &  {\it (0.2101) } 	&-0.088 	&1.49 	& 4.70862         &1.07        & 2.544 	&1.07        & 5.0427 \\
\hline
 \end{tabular}
\caption{
Properties and yields of the H, He and Be isotopes  from ternary  spontaneous fission $^{252}$Cf(sf) relevant for the final distribution of observed H, He nuclei (denoted by the superscript $f$). Experimental  yields $Y^{\rm exp}_{A,Z}$ \cite{Whe67,Wollersheim,Wollersheim1,Jesinger05} are compared to the yields $Y^{\rm interp}_{A,Z}$ obtained from an interpolation formula \cite{Valskii2} as well the yields calculated in different approximations of the nuclear Hamiltonian (\ref{fullH}), see Sections \ref{sec:int}, \ref{sec:med}, together with the corresponding multipliers  $R_{A,Z}(\lambda_T)$, which represent the intrinsic partition functions:  
The final state distribution $Y^{\rm final}_{A,Z}$ considering only the observed stable nuclei $\{A,Z\}^f$,
 the relevant distribution $Y^{\rm rel,\gamma}_{A,Z}$ of noninteracting  nuclei, the relevant distribution $Y^{\rm rel,vir}_{A,Z}$ taking into account continuum correlations, and the relevant distribution $Y^{\rm rel, eff}_{A,Z}$ including interaction between the constituents.  The binding energy $B_{A,Z}$ (in MeV), the degeneracy $g_{A,Z}=2\, J_{A,Z}+1$, and  threshold energy of continuum states $E^{\rm thresh}_{A,Z}$ (in MeV) according Ref. \cite{nuclei}, are also given. The first three rows show the Lagrange parameters $\lambda_i$ obtained for the four different calculations (in  MeV).}
\label{Tab:HHeCf}
\end{center}
\end{table}

In Table~\ref{Tab:HHeCf} we collect some experimental results denoted as $Y^{\rm exp}_{A,Z}$.  For $^1$H and $^2$H, we give the values of  \cite{Whe67}. For $^3$H, $^4$He, $^5$He, $^6$He, $^7$He, $^8$He, we take the values from Refs.  \cite{Wollersheim,Wollersheim1} where also the errors are given. Note that the long-living isotopes  $^6$He and $^8$He which are unstable with respect to $\beta$ decay are stable with respect to strong interaction and are observed in the experiments.
In addition to the final yields $Y^{\rm exp}_{A,Z}$ seen in the experiments, denoted by the superscript $f$ at the stable nucleus, we give also primary yields of short-living, particle unstable nuclei which can contribute to the yields of the observed nuclei. The final yields are related to the primary yields as $Y_{^3{\rm H}^f}=Y_{^3{\rm H}}+Y_{^4{\rm H}},
 Y_{^4{\rm He}^f}=Y_{^4{\rm He}}+Y_{^5{\rm He}}+2 Y_{^8{\rm Be}}, Y_{^6{\rm He}^f}=Y_{^6{\rm He}}+Y_{^7{\rm He}}, Y_{^8{\rm He}^f}=Y_{^8{\rm He}}+Y_{^9{\rm He}}$.
Analyzing the energy spectra, for $^{252}$Cf(sf) the formation of  $^5$He and $^7$He  has been determined \cite{Wollersheim},
and the ratios of primary yields $Y_{^5{\rm He}}/Y_{^4{\rm He}} = 0.21(5)$ and  $Y_{^7{\rm He}}/Y_{^6{\rm He}} = 0.21(5)$ have been reported.
 For $^8$Be, the value 10$\pm 6$ was found for the primary yield in \cite{Jesinger05}.
 
 For $^{252}$Cf the multiplicity of protons emitted in ternary fission is reported to be $6.086 \times 10^{-5}$ \cite{[2]}. This corresponds to an experimental yield $Y^{\rm exp}_{1,1} \approx 190.2$ relative to the yield of $\alpha$-particles.
 Also prompt neutron emission has been measured, for a recent work see \cite{Chietera18}.
The fractional percentage of ternary fission "scission" neutrons has been determined to be $7.6 \pm 2.8 \%$ \cite{Vorobyev,Stuttge}. In Ref. \cite{Stuttge} a temperature of $T_{\rm sci} = 1.2$ MeV has been used.  
A very recent measurement of  the total neutron multiplicity in  $^{252}$Cf fission is $3.81 \pm 0.05$ \cite{Hansell}. This leads to a scission neutron multiplicity of $0.290 \pm 0.008$ that corresponds to an experimental yield $Y^{\rm exp}_{1,0} \approx 0.29/(3.2 10^{-3})\times 10^4=906250$ relative to the yield of $\alpha$-particles.

As pointed out in Ref.  \cite{Mutterer08}, experimental studies  at low energy of ternary-particle-unstable nuclei producing $\alpha$ particles are still scarce, and the data are not very consistent.
The ratio $^6$He$^f$/$^4$He$^f$ was  reconsidered in \cite{Mutterer08}, and a Gaussian fit above 9 MeV energy  gives the value $Y_{^6{\rm He}^f}/Y_{^4{\rm He}^f} =0.041(5)$.

An interpolation formula has been presented in Ref. \cite{Valskii2} and compared to measured data (\cite{Whe67} for H  and  \cite{Dlo92}  for He isotopes). Parameter values for quantities 
similar to the temperature ($\Theta=1.25$ MeV), 
the neutron chemical potential ($\varepsilon_n=2.8$ MeV) and the proton chemical potential ($\varepsilon_p=15.8$ MeV) 
have been fitted, and are also shown in Table~\ref{Tab:HHeCf} together with the yields  $Y^{\rm interp}_{A,Z}$ \cite{Valskii2}, which are calculated with the Valski interpolation formula. To explain the yields observed from ternary fission of $^{252}$Cf(sf), Boltzmann-like distributions are used. These contain  the binding energy $B_{A,Z}$ and the ground-state degeneracy $g_{A,Z}=2\, J_{A,Z}+1$  of nuclei \cite{nuclei} which are also shown in Table~\ref{Tab:HHeCf}. The remaining columns will be discussed in Sections  \ref{sec:int} and \ref{sec:med} below.

\section{Information theoretical description of the distribution of cluster yields}
\label{sec:int}

Information theory considers the problem of reconstructing  a distribution if some information about the ensemble is given. The most probable distribution is 
obtained from the maximum of information entropy if some averages of of the system properties are given. 
From the maximum of information entropy, a Gibbs distribution is obtained with Lagrange parameters 
$\lambda_i$ which are determined self-consistently, by describing the averages.  In this section we consider whether it is possible to reconstruct the distribution of yields
$Y^{\rm exp}_{A,Z}$ with only a limited amount of information about the system. This approach is well-known from equilibrium thermodynamics where the averages of energy and particle numbers
of the conserved components are given to define the grand canonical ensemble, and the Lagrange parameters $\beta=1/T, \mu_n, \mu_p$ occurring in the equilibrium ensemble are related to the temperature and the chemical potentials.

Within the method of the nonequilibrium statistical operator (NSO) \cite{NSO}, at a given time $t$ the relevant statistical operator $\rho_{\rm rel}(t)$ is constructed from 
this maximum entropy principle, and the corresponding Lagrange parameters $\lambda_i(t)$ become functions of time. The statistical operator $\rho(t)$ describing the nonequilibrium evolution of the system follows as 
\begin{equation}
\label{rhoZ}
\rho(t)=\lim_{\epsilon \to 0} \epsilon \int_{-\infty}^t dt' e^{-\epsilon (t-t')}e^{-\frac{i}{\hbar} H (t-t')}\rho_{\rm rel}(t')e^{\frac{i}{\hbar} H (t-t')}
\end{equation}
which solves the Liouville-von Neumann equation, with $H$ the system Hamiltonian given in Eq. (\ref{fullH}). Kinetic equations as well as hydrodynamic equations can be derived within this approach.

We discuss the construction of the relevant statistical operator which is a ingredient to describe the nonequilibrium process. As relevant observables, we consider the averages of neutron number and proton number, as well as the Hamiltonian 
\begin{equation}
\label{fullH}
H=\sum_{\tau,k} E_{\tau}(k) a_{\tau,k}^\dagger a_{\tau,k}+\sum_{\tau,k,k'} V^{\rm ext}_{\tau}(k,k') a_{\tau,k'}^\dagger a_{\tau,k}
+\frac{1}{2}\sum_{\tau,\tau',k,k',p,p'} V^{\rm int}_{\tau,\tau'}(p,k;p',k') a_{\tau',p'}^\dagger a_{\tau,k'}^\dagger a_{\tau,k} a_{\tau',p}
\end{equation}
which describes the interaction of nucleons, $\tau = \{n,p\}$, with an external potential $V^{\rm ext}_{\tau}(k,k')$ and the nucleon-nucleon interaction $V^{\rm int}_{\tau,\tau'}(p,k;p',k')$;
the quantum number $k$ denotes the wave vector and spin of the nucleon with kinetic energy $E_{\tau}(k)$. Averages are going to be calculated with the correspondent relevant operator \begin{equation}
\label{relev}
\rho_{\rm rel}(t)=Z_{\rm rel}^{-1}(t) e^{-[H-\lambda_n(t) N_n -\lambda_p(t) N_p]/\lambda_T(t)}
\end{equation}
where $ Z_{\rm rel}(t) = {\rm Tr} \exp\{-[H-\lambda_n(t) N_n -\lambda_p(t) N_p]/\lambda_T(t)\} $ is the relevant partition function, $N_\tau$ denotes the particle number of neutrons/protons, and the Lagrange multipliers are going to be eliminated by the known informations, such as internal energy and particle densities of the system. 

The solution of this many-particle problem is not simple and needs some approximations, such as replacing the Hamiltonian by a more simple model which can be solved. Such simple model Hamiltonians are the ideal nucleon gas or the mean-field approximation, where the Hamiltonian describes a noninteracting system of quasiparticles. We are interested in the formation of bound states so that,
in a first approximation, we consider the ideal energy functional 
\begin{equation} 
\label{H0}
H^{(0)} = \sum_{A,Z,{\bf P}} g_{A,Z} \left(-B_{A,Z}+\frac{\hbar^2 P^2}{2 A m}\right)
\end{equation} 
with $\bf P$ the center of mass momentum for the cluster $\{A,Z\}$, $m$ is the average nucleon mass. This model Hamiltonian describes the nucleon system as
an ideal mixture of non-interacting free nucleons and nuclei.
We allow for reactions between the different components $\{A,Z\}$ so that  the number of each component is not conserved  but only 
the total number of neutrons and protons in the system. This approximation can be applied in the low-density case where the interaction between the constituents of the nuclear system becomes weak.
The Lagrange parameters in (\ref{relev}) calculated to reproduce the observed distribution with the model Hamiltonian $H^{(0)}$ (\ref{H0}), are denoted by $\lambda^{(0)}_T,\lambda^{(0)}_n,\lambda^{(0)}_p$, respectively.

The problem to eliminate the Lagrange multipliers by the given averages of internal energy and particle number densities is well known from statistical physics and leads to the Fermi/Bose distribution. Before we discuss the corresponding results,
we emphasize that initially we are discussing only the parametrization of the measured yields, using information about the observed nuclei such as ground state binding energy and degeneracy. 

The obtained Lagrange  parameters  $\lambda^{(0)}_i$ should not be interpreted as thermodynamic quantities like temperature and chemical potentials for 
several reasons:\\
 (i) The energy functional is only an approximation. The full energy functional should also include excited states and interactions. The full information about the Hamiltonian of the system leads to the quantum statistical approach.\\
(ii)  Fission is a nonequilibrium process and is not described by an equilibrium distribution. 
We have to include also the dynamical information which is described by the full Hamiltonian 
and contains the information on the final distribution measured in the experiment as well as information of the distribution in the past $t' \le t$, see Eq. (\ref{rhoZ}). 
This leads to the so-called relevant distribution $Y^{\rm rel}_{A,Z}(t)$ which depends on the time $t$,
and the evolution of the distribution with time described by generalized reaction kinetics \cite{NSO}.
We will not discuss how the evolution of the system follows from the solution of (\ref{rhoZ}) 
 but use instead the simple concept of the freeze-out approximation.
This means that the relevant distribution is given by thermodynamic equilibrium up to the freeze-out time. After that,
the relevant distribution evolves according to reaction kinetics as described by the decay of excited states. 
The different versions of $Y^{\rm rel}_{A,Z}$ shown in Table~\ref{Tab:HHeCf} correspond to different model Hamiltonians as an approximation to the nucleon Hamiltonian (\ref{fullH}).\\
(iii)  The system to be described is not homogeneous nuclear matter as in thermodynamic equilibrium but is inhomogeneous. 
A local density approximation is problematic. In addition, the relation between the Lagrange parameters $\lambda_i$ and the thermodynamic quantities,
in particular the densities $n_n({\bf r},t), n_p({\bf r},t)$, is not the relation between temperature, chemical potentials and density 
as known from non-interacting, ideal quantum gases but is more complex. Our aim is to find arguments to infer the density from the data.

The information of the properties of the observed nuclei, see Tab.  \ref{Tab:HHeCf}, leads to the Boltzmann-like distribution (nondegenerate case)
\begin{equation} 
\label{Y0}
Y^{(0)}_{A,Z} \propto  n^{(0)}_{A,Z}=g_{A,Z} \left(\frac{2 \pi \hbar^2}{A m \lambda^{(0)}_T}\right)^{-3/2}
e^{(B_{A,Z}+(A-Z) \lambda^{(0)}_n+Z  \lambda^{(0)}_p)/ \lambda^{(0)}_T}.
\end{equation} 
Because we don't know the prefactor we consider only the ratio $Y_{A,Z;\alpha}=Y_{A,Z}/Y_{^4{\rm He}^f}\times 10^4$, i.e., the isotope yield relative to the final yield of $\alpha$ particles.
Following convention, the final yield of the $\alpha$ particles is fixed as $Y[^4{\rm He}^f]=10000$.

We infer the Lagrange parameter values  $\lambda^{(0)}_i$ based on the information about the observed final nuclei 
by minimizing the sum over the relative square deviation $(Y^{(0)}_{A,Z;\alpha}-Y_{A,Z;\alpha})^2/Y_{A,Z;\alpha}$. 
For instance, considering the four lightest and most abundant isotopes, i.e. the yield ratios of  $^3$H$^f$, $^4$He$^f$, $^6$He$^f$, and $^8$He$^f$, we determine 
 the values of the three Lagrange parameters
 $\lambda^{\rm final}_T=0.806219$ MeV, \,\,\,\, $\lambda^{\rm final}_n=-2.19217$ MeV, \,\,\,\, $\lambda^{\rm final}_p=-17.7023$ MeV. 
 The values for the corresponding yields $Y^{\rm final}_{A,Z}$ (we drop the index $\alpha$) is shown in Table~\ref{Tab:HHeCf}.

This approach is similar to that used in the Albergo determination of temperature and chemical potentials from the observed yields of nucleons and nuclei 
from heavy ion collisions \cite{Albergo}. However, we cannot identify the Lagrange  parameters $\lambda^{(0)}_i$ with the thermodynamic parameters temperature and chemical potential 
of the nuclear system because, and as stated above,
the ideal energy functional  (\ref{H0}) considers only the ground states of the nuclei, and the interaction between the nucleons/nuclei is neglected. 
In addition, the correct nonequilibrium distribution is given by $\rho(t)$ (\ref{rhoZ}) which coincides with $\rho_{\rm rel}(t)$ only in thermodynamic equilibrium.

Evidently, our first approach which relates the final distribution to the binding energies of nuclei and their degeneracies is not sufficient. 
There are further unstable nuclei and excited states which should be taken into account. In addition to the stable isotopes $^1$H, $^2$H, $^3$He, $^4$He we have 
$^6$He and $^8$He which are unstable with respect to weak processes ($\beta$-decay) but have a sufficiently long half-life so that they are observed like 
stable nuclei. Other isotopes such as $^4$H, $^5$He, $^7$He, $^9$He, $^8$Be are unstable with respect to the strong interaction and decay immediately 
so that these primary nuclei are not detected as final yields. They should appear in the relevant distribution if the model Hamiltonian (\ref{H0}) contains the sum over all bound states.
We have calculated the expected yields $Y^{\rm final}_{A,Z}$ of the unstable isotopes using the Lagrange parameters $\lambda^{\rm final}_i$. These are given 
in Tab.   \ref{Tab:HHeCf}, col. $Y^{\rm final}_{A,Z}$, in italic parentheses. We conclude that an essential part of clustering is found within these unstable nuclei. 

This problem, that also the formation of unstable nuclei are expected, is solved by taking into account that the observed distribution are not equilibrium distributions but the result of a time evolution described by $\rho(t)$ (\ref{rhoZ}).
To solve this in a simple approximation, we assume a freeze-out scenario. Up to the  freeze-out time $t'=t_{\rm freeze}$, 
only the information about energy density and particle number density is sufficient to describe the state of the system. 
The relevant distribution (\ref{relev}) can be used to describe the system.

After this, a more detailed description of the system is necessary where the occupation numbers of quasiparticle states of the components are relevant, the corresponding Lagrange parameters are the distribution functions. This stage of evolution is described by reaction kinetics, unstable nuclei decay that feed the states of observed stable nuclei.
The primary distribution, described by the relevant statistical operator $\rho_{\rm rel}(t_{\rm freeze})$ and the yields $Y_{A,Z}^{\rm rel}$, is transformed to the yields of detected 
nuclei $Y_{A,Z^f}$ denoted as feeding in Section \ref{sec:exp}, for instance $Y_{^3{\rm H}^f}=Y_{^3{\rm H}}+Y_{^4{\rm H}}$, etc.

%

%

Another consideration is the inclusion of  excited states of nuclei to 
characterize the relevant distribution, see the data tables in Ref. \cite{nuclei}. Excited states contribute to the statistical weight 
of an isotope. For instance, the isotope $^4$H has an excited state at 0.31 MeV with a degeneracy factor 3. Assuming that this excited state is also populated at freeze-out
described by the relevant distribution, it decays and its yield will be found in the corresponding final cluster state. 
The statistical weight or intrinsic partition function of a special channel 
characterized by   $\{A, Z\}$ contains not only the bound states but also continuum correlations, see \cite{R20}. 
The threshold energy $E^{\rm thresh}_{A,Z}$ denotes the edge of continuum states and it is also shown in Tab. \ref{Tab:HHeCf}. 
In general, it is given by the neutron separation energy $S_n$, but in some cases other decay channels such as $S_{2n}$ ($^6$He,  $^8$He) or $\alpha$-decay ($^8$Be) 
determine the edge of continuum states. At present, we neglect the contribution of continuum correlations, but will discuss them below in Section \ref{sec:cont}.
The corresponding distribution is denoted as $Y^{\rm rel, \gamma}_{A,Z}$.

The account of excited states can be realized introducing in Eq. (\ref{Y0}), which considers only the ground state with lowest energy, the prefactor $R^\gamma_{A,Z}(\lambda_T)$,
\begin{equation}
\label{Rgamma}
R^\gamma_{A,Z}(\lambda_T)=1+\sum^{\rm exc}_i \frac{g_{A,Z,i}}{g_{A,Z}} e^{-E_{A,Z,i}/\lambda_T},
\end{equation}
which is related to the intrinsic partition function, so that $Y^\gamma_{A,Z}=R^\gamma_{A,Z} Y^{(0)}_{A,Z}$.
The summation is performed over all excited states, excitation energy $E_{A,Z,i}$ and degeneracy $g_{A,Z,i}$  \cite{nuclei} , which decay to the ground state.
The result $Y^{\rm rel, \gamma}_{A,Z}$ shown in Tab. \ref{Tab:HHeCf} was obtained with the factor $R^\gamma_{4,1}(\lambda_T)=1+3/5 e^{-0.31/\lambda_T}$ for $^4$H, 
and  $R^\gamma_{8,4}(\lambda_T)=1+5 e^{-3.03/\lambda_T}$ for $^8$Be.
No excited states below the continuum edge are known for the other bound nuclei so that the remaining factors $R^\gamma_{A,Z}(T)$ are unity. 
Assuming that the unstable nucleus $^4$H feeds  the measured yield of $^3$H$^f$, that $^5$He and $^8$Be feed $^4$He$^f$,
 $^7$He feeds $^6$He$^f$, 
and $^9$He feeds $^8$He$^f$, the optimization with respect to the measured yields  $Y^{\rm exp}_{A,Z}$ using the least squares  method gives  the values of the three Lagrange parameters
 $\lambda^{\rm rel, \gamma}_T=1.28018$ MeV, \,\,\,\, $\lambda^{\rm rel,\gamma}_n=-3.18741$ MeV, and $\lambda^{\rm rel,\gamma}_p=-16.5882$ MeV. 
 
 The inferred yields $Y^{\rm rel, \gamma}_{A,Z}$ for the final distribution reproduce nicely the measured values $Y^{\rm exp}_{A,Z}$ for $^2$H, $^3$H$^f$, and $^6$He$^f$ in relation to $^4$He$^f$. It seems that $^6$He$^f$ is overestimated, and $^8$He$^f$ is underestimated. Notably, the  yields of the unstable nuclei $^5$He and $^7$He which have been inferred from the energy spectra of emitted
 $\alpha$ particles  \cite{Wollersheim} are also overestimated by the relevant distribution $Y^{\rm rel, \gamma}_{A,Z}$. The prompt emission of protons and neutrons  will be discussed below.

\section{Relevant distribution derived from the  full Hamiltonian}
\label{sec:med}

The estimate $Y^{\rm rel, 0}_{A,Z}$ should be improved taking into account different effects to be discussed in the following.\\ (i) It is not consistent to consider only the 
excited bound states below the edge of continuum states and to neglect correlations in the continuum. In particular,  $^4$H, $^5$He, $^7$He, $^9$He, and $^8$Be are not bound
but appear as correlations in the continuum. Continuum correlations should be also considered for other, weakly bound nuclei such as $^2$H, $^6$He. We need a systematic treatment of 
the contribution of continuum correlations. This is possible with the help of the generalized Beth-Uhlenbeck formula \cite{BU}. 
The corresponding relations are denoted as virial equations.\\
(ii) Instead of the approximation (\ref{H0}) where interaction between nucleons and nuclei is neglected, we have to consider in-medium effects if we treat the full Hamiltonian (\ref{fullH}).\\
(iii) The nuclear system is not homogeneous. We should consider the mean field near the scission point where both main fragments are close together, 
and the nucleons forming the neck region are not correctly described by a homogeneous gas. This means that the full Hamiltonian contains in addition to the interaction between
the constituents also the mean field $V^{\rm ext}_{\tau}(k,k')$ of the main fragments of scission.

\subsection{Continuum correlations}
\label{sec:cont}

A comprehensive discussion of the contribution of continuum correlations in the case of $^4$H, $^5$He has been given in \cite{R20} based on the generalized Beth-Uhlenbeck formula.
For $^2$H extended discussions have been given earlier, see references given in  \cite{R20}, and for $^8$Be, see also \cite{HS}. The model calculations are compared to
measured phase shift data. The account of continuum correlations leads to the virial expansion, which is represented by the reduction factor $R_{A,Z}^{\rm vir}(\lambda_T)$.
The virial expansion for the deuteron channel $d$ ($A=2, Z=1$) is obtained from the Beth-Uhlenbeck expression
\begin{equation}
\label{BU}
R_{d}^{\rm vir}(\lambda_T)= 1-e^{-E^{\rm thresh}_d/\lambda_T} +e^{-E^{\rm thresh}_d/\lambda_T}\frac{1}{\pi \lambda_T} \int_0^\infty dE e^{-E/\lambda_T}\delta_d(E)
\end{equation}
where $E$ denotes the c.m. energy of the $n - p$ system describing the correlations of the deuteron channel above the continuum edge. The scattering phase shifts are denoted as
$\delta_d(E)$. From the known values of these scattering phase shifts, see \cite{HS}, we have for instance the value $R_{d}^{\rm vir}(1.3\, {\rm MeV})= 0.98$, see Appendix \ref{app1}. 

The $^4$H channel is treated similarly. 
Because there is no bound state, only the last term in (\ref{BU}) with the integral over the scattering phase shifts in the $t - n$ channel remains.
The corresponding virial coefficient in the $t - n$ channel has been calculated in \cite{R20} and parametrized introducing an effective energy $E^{\rm eff}_{tn}(T,n_n)$ so that
the value of the reduction factor 
\begin{equation}
\label{Rvir}
R_{4,1}^{\rm vir}(\lambda_T)=e^{-E^{\rm eff}_{t,n}(\lambda_T,0)/\lambda_T+E^{\rm thresh}_{4,1}/\lambda_T}
\end{equation}
 follows as 
$R_{4,1}^{\rm vir}(1.3\,{\rm MeV})=0.06064$. 

For the $^5$He channel, the virial coefficient using the $\alpha - n$ scattering phase shifts is calculated, see \cite{HS,R20}.
For the reduction factor of $^5$He we find from the relation similar to (\ref{Rvir}), and using the parametrization $E^{\rm eff}_{\alpha,n}(\lambda_T,0)$ given in \cite{R20}, 
the value $R_{5,2}^{\rm vir}(1.3\, {\rm MeV})=0.70441$. For $^8$Be, the virial coefficient using the $\alpha - \alpha$ scattering phase shifts is calculated, see \cite{HS,R20}.
 
It is not easy to calculate the continuum correlations for arbitrary clusters $\{A, Z\}$. 
It seems that continuum correlations are important for weakly bound states, so that the edge of the continuum $E^{\rm thresh}_{A,Z}$ is of relevance.
Therefore we introduce an interpolation formula, in units of MeV,
\begin{equation}
\label{fitRvir}
R_{A,Z}^{\rm vir}(\lambda_T)=1/(e^{-(E^{\rm thresh}_{A,Z}+1.12949)/0.204007}+1)/(e^{-(E^{\rm thresh}_{A,Z}+2.44619)/\lambda_T}+1)
\end{equation}
which reproduces the values given above for $^2$H, $^4$H, and $^5$He at $\lambda_T=1.3$ MeV.

We used this interpolation formula to infer the reduction factor $R_{A,Z}^{\rm vir}(\lambda_T)$ for the remaining isotopes as given in Tab. \ref{Tab:HHeCf}.
It replaces $R_{A,Z}^{\gamma}$ (\ref{Rgamma}) so that $Y^{{\rm rel, vir}}_{A,Z}=R_{A,Z}^{\rm vir} Y^{(0)}_{A,Z}$.
The measured yields $Y^{\rm exp}_{A,Z}$ are better reproduced, in particular the results for $^4$H,  $^5$He and $^7$He are significantly modified. 
However, even with the account of continuum correlations  the yield of $^6$He is overestimated as before, and $^8$He and $^8$Be remain underestimated. We have to consider additional effects 
which are of relevance to calculate the relevant distribution.\\

\subsection{Pauli blocking}

Another interesting problem is the medium modifications, that should be taken into account when considering a quantum statistical treatment of the Hamiltonian (\ref{fullH}) of the nuclear system, treating the interaction between the components. In lowest order, we have self-energy shift and Pauli blocking, see \cite{R20} and references given there.
If we neglect the momentum dependence of the single-nucleon self-energy shift (rigid shift approximation), the self-energy shifts of bound and scattering states 
are identical so that the binding energy is not changed. Then, it can be incorporated in a shift of the parameters $\lambda_n,  \lambda_p$. 
The Pauli blocking leads to a decrease of the binding energy (we denote the disappearance of the binding energy as the Mott effect). 
Because it is related to the occupation of single-particle nucleon states by the medium, it is sensitive to the nucleon densities $n_n, n_p$.

The effect of Pauli blocking is expected to lead  to dissolution of weakly bound states which are shifted to the continuum. The Pauli blocking has been considered for 
the bound states in former publications, see \cite{R20}, where also the effect of Pauli blocking for $^4$H and $^5$He is calculated. 
To evaluate the Pauli blocking shifts we use the results given in \cite{R20}. For the unstable nuclei $^4$H, $^5$He we use the Pauli blocking
shifts of the constituent cluster $t, \alpha$, respectively, and the density-dependent contribution of scattering phase shifts according to the generalized Beth-Uhlenbeck formula.

Now we infer the effective reduction factor $R_{A,Z}^{\rm eff}(\lambda_T)$ for the different components $\{A,Z\}$ of the relevant distribution, which are needed to reproduce 
the observed yields,
and suppose that density effects are responsible for these inferred values.
Because density effects become more visible for weakly bound systems, we consider the reduction factor $R_{6,2}^{\rm eff}(\lambda_T)$ for $^6$He 
as an unknown  quantity to be determined from the fitting of the parameters (the threshold energy for the 
continuum states 0.975 MeV is small compared also to the deuteron case where it amounts to 2.225 MeV). The reduction factors for $^2$H, $^3$H, $^4$H, $^3$He, $^4$He, 
$^8$He, $^9$He, and $^8$Be remain unchanged and coincide with $R_{A,Z}^{\rm vir}(\lambda_T)$, see Tab. \ref{Tab:HHeCf}, but for $^5$He, $^6$He, and $^7$He, 
 $R_{A,Z}^{\rm eff}$ is very different from  $R_{A,Z}^{\rm vir}$. 
We know the yields of $^5$He, $Y_{5,2}/Y_{4,2} = 0.21$ \cite{Wollersheim} and the yield of   $^7$He, $Y_{7,2}/Y_{6,2} = 0.21$ \cite{Wollersheim}.   
We also know the final yield of $^6$He which is given in Ref.  \cite{Wollersheim} as 270, see Tab.  \ref{Tab:HHeCf}. We can infer the effective reduction factors 
$R_{A,Z}^{\rm eff}(\lambda_T)$ that reproduce these values. The values shown in  Tab.  \ref{Tab:HHeCf} are constructed this way. They are lower than the 
values $R_{A,Z}^{\rm vir}(\lambda_T)$. This means that $Y^{\rm rel,eff}_{A,Z}$ for $^5$He, $^6$He, and $^7$He was calculated by using the experimental values of the yields to calculate $R^{\rm eff}_{A,Z}$ and then with that, we calculated $Y^{\rm rel,eff}_{A,Z} = Y^{\rm exp}_{A,Z}$ for these three isotopes. Yields for the isotopes $^4$H, $^3$He are not observed
so that we have only the predictions $Y^{\rm rel,eff}_{A,Z} $ for these isotopes, also for $^8$Be we give only predictions because the errors are quite large.

We remember that taking into account only the virial expansion, i.e., the contribution of continuum correlations, $^6$He is overestimated, $^8$He is underestimated.
A possible explanation may be the small binding energy 0.975 MeV below the continuum edge for $^6$He which makes this state more sensitive to medium shifts 
so that the Pauli blocking is stronger. Note that an alternative explanation could be the formation of tetraneutron correlations in neutron matter \cite{tetra}. 
In the effective approximation  $Y^{\rm rel, eff}_{A,Z}$ only $^2$H, $^3$H, $^4$He, and $^8$He have been used to determine the Lagrange parameters, 
because the threshold energy for the continuum is larger. The measured contributions of $^5$He, $^7$He have been adopted, values for $^9$He are assumed. The multipliers  $R_{A,Z}^{\rm eff}(1.3)$ are determined from the measured yields of $^5$He, $^6$He, $^7$He. 
It is found that the inferred values $R_{A,Z}^{\rm eff}(1.3)$ are smaller than the values $R_{A,Z}^{\rm vir}(1.3)$ for these isotopes.

We conclude that the result $Y^{\rm rel, vir}_{A,Z}$, in which continuum correlations are taken into account as virial coefficients, but medium corrections are neglected, has some deficits.\\
i) The yield of $^6$He and $^7$He is overestimated. Because of the weak binding of these isotopes, compared to others including $^8$He, the dissolution of the bound state in dense matter may be of relevance. Instead of a calculated  reduction factor  $R_{6,2}^{\rm vir}(1.3)=0.9453$ for $^6$He, the data give the observed reduction factor $R_{6,2}^{\rm eff}(1.3)=0.696083$.\\
ii) The yield of $^5$He, $^7$He is also overestimated. For $^5$He, instead of the calculated reduction factor $R_{5,2}^{\rm vir}(1.3)= 0.7044$ 
the value  $R_{5,2}^{\rm eff}(1.3)=0.659566$  follows from the data. 
For $^7$He, the calculated virial value $R_{7,2}^{\rm vir}(1.3)=0.821$ for the reduction factor is replaced by the empirical value $R_{7,2}^{\rm eff}(1.3)=0.398762$, see Tab. \ref{Tab:HHeCf}. Note that the values have large error bars.

Having calculated the empirical values $R_{A,Z}^{\rm eff}$ of the reduction factor of the light nuclei, after the subtraction of the effect of continuum correlations,
the remaining part contains the in-medium effects. From this, we estimate the neutron density of the medium at freeze-out time.
From the $^5$He values and the results given in the paper \cite{R20}, the value $n_n=0.000254$ fm$^{-3}$ is obtained if a relation similar to Eq. (\ref{Rvir}) is used. 
The density dependence of $E^{\rm eff}_{\alpha,n}(\lambda_T,n_n)$ is given in \cite{R20}.

For $^6$He, the inferred reduction factor $R_{6,2}^{\rm eff}(1.3)=0.696083$ inserted in Eq. (\ref{fitRvir}) gives a shift of $E$ of about 1.7 MeV. The Pauli blocking shift according to 
\cite{R20} is $4 \times 532.0 e^{-0.051 T} n_n$ (separation of two neutrons, $F_{A,Z} \approx 2$)
so that the value $n_n=0.0007988$ fm$^{-3}$ is obtained.
The reduction factor for $^7$He is about 0.5. Such values are obtained for $^4$H at a density of about $n_n=0.002$ fm$^{-3}$. 
Note that the measured values have large errors, and also the treatment of medium effects should be improved.

A more recent analysis of experimental data has been performed in Ref. \cite{Mutterer08}. Instead of the ratio $^6$He/$^4$He = 0.031(2) given in \cite{Wollersheim}, 
a value 0.041(5) has been presented. A higher value of $^6$He/$^4$He would also give a higher value of the  reduction factor $R_{6,2}^{\rm eff}(1.3)=0.920626$ which corresponds to the $E=-0.50$ MeV according to Eq. (\ref{fitRvir}) so that the Pauli blocking shift is only 0.475 MeV. 
The corresponding density  follows as $n_n=0.000223$ fm$^{-3}$ what is in better agreement with the former result.

The measured value $Y_{8,4}=10(6)$ for $^8$Be \cite{Jesinger05} also underlines the better fit using the  distribution $Y^{\rm rel, eff}_{A,Z}$.

\section{External mean-field potential}
\label{sec:LDA}

A further improvement of the treatment of the Hamiltonian of the nucleon system is to take into account the interaction with the two large fragments after scission.
This can be done introducing a mean field, produced by the strong nucleon-nucleon interaction as well as by the Coulomb interaction.

The neck region where clusters are formed is influenced by the larger fission fragments. There is the Coulomb field which determines the kinetic energy of the emitted particles, but also the strong interaction described, 
e.g., by the pion-exchange potential. These interactions should be added to the Hamiltonian as external fields $ V^{\rm ext}_{\tau}$, Eq. (\ref{fullH}).
Obviously, the use of results such as the yields of the isotopes which are obtained for homogeneous systems, is only possible in the local density approximation (LDA), but demands further discussions. The bound state clusters are compact objects, the wave functions are extended over a region of some fm. A local density approximation may be possible.
In contrast, the Fermi wave number corresponding to baryon density $n_B$ follows as $k_F=(3 \pi^2n_B/4)^{1/3}$ 
(symmetric matter). For $n_B = 0.0004$ fm$^{-3}$, the value $k_F=0.144$ fm$^{-1}$ follows. The wave function
of neutrons is rather extended so that a LDA approach is not justified.

We assume that the relevant fragment distribution, including the two large fragments as well as the light clusters or correlations, are already formed at the scission point, as also 
known from the scission point yield (SPY) model, see, e.g., 
\cite{Lemaitre,Bertsch,Karthika,Bulgac,Carjan} for recent work.
Hartree-Fock-Bogoliubov and related mean-field calculations have been performed for fission. To describe cluster formation, one has to go beyond 
a mean-field approach. A similar problem arises when describing the $\alpha$ decay of heavy nuclei where a 
quartetting wave function approach has been proposed to describe the preformation of $\alpha$-like correlations \cite{Xu}.

As a simple estimate we consider the 
superposition of two Woods-Saxon potentials at distance $R_1+R_2+d$. For simplicity we assume the symmetric case
where $^{252}$Cf decays into two fragments approximately $^{124}_{48}$Cd (in general, asymmetric decays occur. 
Experimentally, for $^{252}$Cf, the highest
yield values were found for $^4$He + $^{101}$Zr + $^{147}$Ba, see \cite{Karthika}) and calculate the Woods-Saxon potentials according to \cite{WoodsS} (units: fm, MeV)
\begin{equation}
\label{WSwf}
 V^{\rm mf,WS}_{n/p}(r )= -52.06 \frac{ 1 \mp 0.639 (N-Z)/A}{1+e^{(r-1.26 A^{1/3})/0.662}}
\end{equation}
for neutrons (upper sign). For protons (lower sign) we have to add the Coulomb potential 
\begin{equation}
\label{WSwfC}
 V^{\rm Coul}(r )= Z e^2 \left\{\frac{3 R_A^2-r^2}{2R_A^3}, \qquad r \le R_A, \qquad \frac{1}{r}, \qquad r > R_A\right\}
\end{equation}
with $R_A = 1.26 A^{1/3}$ fm \cite{WoodsS}.
\begin{figure}[h] 
\centerline{\includegraphics[width=300pt,angle=0]{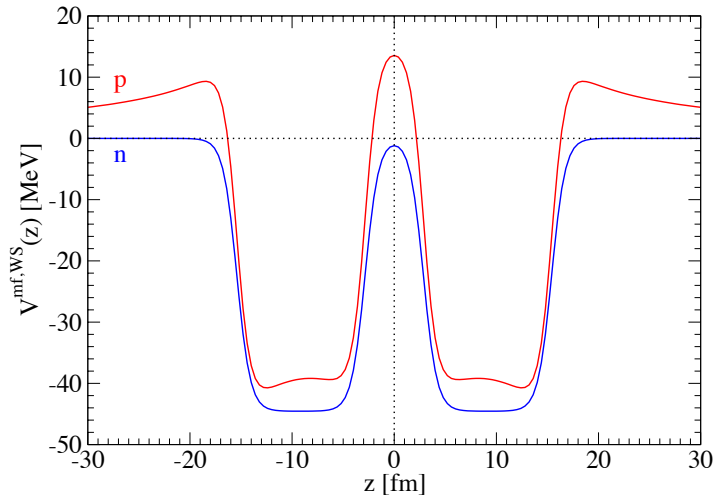}}
\caption{Woods-Saxon potentials of two nuclei $^{124}_{48}$Cd at distance $2 R_A+d$, $R_A=6.28$ fm, $d=5.7$ fm.
Neutrons (blue) and protons (red).}
\label{fig:mf}
\end{figure}
The parameter $d$ for the scission point is under discussion, 
a recent estimate \cite{Panebianco} gives the range 4 - 6.5 fm and a proposed value of 5.7 fm.
The distance between the fragment center of mass points is $2 R_A+d=18.06$ fm. The mean-field (MF) potentials for neutrons and protons along the symmetry line $z$ is shown in Fig \ref{fig:mf}. The values at the scission point $z=0$ are $V^{\rm mf,WS}_n(0)=-1.1867$ MeV, $V^{\rm mf,WS}_p(0)+ V^{\rm Coul}(0)=13.549$ MeV. Larger estimates of the scission parameter $d$ are reported for ternary fission  with larger clusters such as $^{50}$Ca \cite{Dubna} which will not be discussed here.

It is possible to estimate the neutron density at the scission point and to understand whether the density values derived from the Pauli blocking estimates are reasonable.
 For an exploratory calculation, we use the  parametrization of neutron/proton densities in nuclei
 \begin{eqnarray}
&& n_{n/p}( r)=\frac{n_{0,n/p}}{1+e^{(r-R_{n/p})/a_{n/p}}}
\end{eqnarray}
with $R_{n}=(0.953 N^{1/3}+0.015 Z+0.774)$ fm, \,\,$R_{p}=(1.322 Z^{1/3}+0.007 N+0.0224)$ fm, and diffusivities
$a_{n}=(0.446 +0.072 N/Z)$ fm,\,\, 
$a_{p}=(0.449 +0.071 Z/N)$ fm  \cite{Seif}.
The neutron density at distance $d/2=2.85$ fm from the surface is  $n_n=0.00059648$ fm$^{-3}$ for each $^{124}_{48}$Cd nucleus so that the value of the neutron density in the neck region is about $n_n(z=0)=n^{\rm scission}_n\approx 0.0012$ fm$^{-3}$. The proton density at this distance is estimated as $n_p(z=0)= n^{\rm scission}_p\approx 0.000315$ fm$^{-3}$. If the distance $d$ for scission is larger,  $d/2=4$ fm, the corresponding densities are smaller, $n^{\rm scission}_n\approx 0.000154$ fm$^{-3},\,\,n^{\rm scission}_p\approx 0.000029$ fm$^{-3}$. 

At this point we emphasize that the treatment of the full Hamiltonian including the external mean-field potential leads to the result that the cluster distributions, 
including the formation of light nuclei, happens at scales of the rms radii which are of the order 1 - 3 fm. The center-of-mass (c.m.) motion of the cluster is determined by the external potential, 
whereas the intrinsic 
properties are determined locally. Intrinsic energy, but also excitations of nuclei and their  distribution, are determined 
by the local properties. The wave function of the c.m. motion is also extended but may be approximated 
by a quasi-classical approximation. This is not possible for the protons and neutrons which are described by extended states. We cannot treat them like plane waves describing free particles but have to use, e.g., quasiparticle 
states in the mean-field potential known from HFB calculations.

As a consequence, single-particle modifications of the in-medium few-particle Schr{\"o}dinger equation
 like the Hartree-Fock self-energy or the Pauli blocking should be expressed in terms of the quasiparticle wave function
 and the occupation numbers of these quasiparticle states. It is possible to introduce Wigner functions and to perform a local approximation, but it has to be noted that Pauli blocking and exchange interaction are nonlocal.
 
 For the yields we conclude that all bound states of the light isotopes  may be described approximatively by local parameters, 
 in particular the local density approximation with the mean field at the scission point $z=0$. The corresponding relevant distribution evolves to the final yields and unstable states feed the corresponding  stable nuclei seen 
 in the experiment. The distribution of neutrons and protons is not described by a local density approximation 
 but by quasiparticle states obtained from the  HFB calculation or other approaches to solve the in-medium
 Schr{\"o}dinger equation for a nucleon moving in an external, mean-field potential. 
 For the  relevant distribution, the neck densities $n^{\rm rel}_n, n^{\rm rel}_p$ at the scission point appear as 
 new parameters. The partial densities $n_n^{\rm rel}, n_p^{\rm rel}$ are not described by the Lagrange parameter
 $\lambda_i$ as in a  local density distribution, but need the solution of the Schr{\"o}dinger equation with an 
 position dependent external potential. The relevant distribution of quasiparticle states evolve also to the
 final distribution which is not described here, but have been discussed extensively in the literature (e.g., Boltzmann equations). 
 At the moment, we  consider $n_n^{\rm rel}, n_p^{\rm rel}$ as additional parameters characterizing the distribution of clusters.

\section{Conclusions: Relation to thermodynamics}
\label{sec:therm}

The ternary fission, considered in this work, is a nonequilibrium process. A fundamental approach, for instance the method of the nonequilibrium statistical operator \cite{NSO},
is required for a systematic treatment. An indispensable requisite is the introduction of the relevant statistical operator reflecting the informations we have about the 
evolving system. In heavy-ion collisions the hot and dense nuclear matter evolves like a fireball, described by local thermodynamic equilibrium. 
In the case of spontaneous fission, the concept of a fireball is hard to accept if one assumes that the nucleus before fission is described as a pure quantum state.
The freeze-out concept is very successful in describing the formation of clusters and the measured yields, 
but appears presently as a semi-empirical approach which requires more sophisticated reasoning within a fundamental nonequilibrium approach.
We use here a information-theoretical approach which does not require the concept of equilibrium.

We have introduced Lagrange parameters $\lambda_i$ to characterize the state of the system, in particular the distribution function for the different components of nuclear matter.
Only in thermodynamic equilibrium are these parameters equivalent to the thermodynamic quantities $T, \mu_n, \mu_p$. 
This is not valid in the fission process considered here. We have a nonequilibrium situation. 
However,  the Lagrange parameter $\lambda_i$ may be considered as the nonequilibrium generalization of these 
thermodynamic parameters. For instance, the large difference $\lambda_n-\lambda_p \approx 13.6$ MeV may be compared with the difference 12.4 MeV of the mean-field potentials shown in Fig. \ref{fig:mf} at $z=0$.

The neck region of the fission process is not homogeneous but exhibits strong gradients in the density distribution, the nuclear mean-field potential and the Coulomb potential.
This has to be taken into account if statistical models are used to explain the observed yields. 
In our approach, the construction of the relevant statistical operator $\rho_{\rm rel}$ has to be improved considering the full Hamiltonian, 
which contains the position dependent external potential. A mean-field (Hartree-Fock) calculation can provide us with more realistic single-particle orbitals.
A local-density approximation where in equilibrium $\lambda_\tau$ is interpreted as chemical potential $\mu_\tau$ 
to calculate the neutron/proton density within the ideal Fermi gas model is not valid. In contrast to the cluster states which are localized in the range of few fm, 
the single-nucleon states are extended and have to be calculated for the mean-field potential $V^{\rm mf}_\tau({\bf r})$ as function of the position $\bf r$.

Within this work, we have  discussed the information-theoretical approach which leads, in the simplest approximation, to a Boltzmann-like distribution. We improved the treatment of the Hamiltonian
of the nucleon system taking the formation of excited, unstable states into account, as well as their decay. In addition, we considered correlations in the continuum and in-medium 
effects. This way we put forward a quantum statistical treatment of the many-nucleon system. Obviously, the excited states and the resonances should be treated in a manner similar
to the stable bound states observed in the final distribution. The inclusion of correlations in the continuum has to be considered, for instance, 
using the scattering phase shifts as shown by the Beth-Uhlenbeck formula. This improves the treatment of very short-living excitations,
but such broad resonances cannot be treated like stable, well-defined bound states as done  in a simple nuclear statistical equilibrium calculation.

Of particular importance are in-medium modifications which lead to a modification of the quasiparticle energy and possibly the dissolution of bound states (Mott effect).
This is of special significance for the weakly bound, neutron-rich exotic nuclei which are strongly influenced by the medium. They may be used as a sonde to probe the environment.
The strong reduction of the yield of the exotic nuclei, observed in many ternary fission experiments, may be explained as a density effect.  
It gives information about the state of the nucleon system at the time instant where the 
chemical composition freezes out.

Further considerations such as the formation of heavier nuclei like droplet condensation \cite{Sarah} have to be included, treating nucleation as a nonequilibrium process.
Future work, in particular the treatment of the external mean-field potential, may give a more detailed description of the nonequilibrium properties and the evolution of cluster distribution in ternary fission processes.\\

{\bf Acknowledgment:} This work was supported by the German Research Foundation (DFG), Grant  \# RO905/38-1, 
the United States Department of Energy under Grant \# DE-FG03-93ER40773, the FCT (Portugal) Projects No. UID/FIS/04564/2019 and UID/FIS/04564/2020, and POCI-01-0145-FEDER-029912, and by PHAROS COST Action CA16214. H.P. acknowledges the grant CEECIND/03092/2017 (FCT, Portugal).

\appendix
\section{Virial expansion}
\label{app1}

Instead of considering unbound systems (resonances) as bound states, we can calculate their contribution to the density (continuum correlation) according to the Beth-Uhlenbeck formula. In particular, for $^2$H, deuteron channel,
we have (c.f. \cite{BU,HS})
\begin{equation}
b_{pn}(T)=\frac{3}{2^{1/2}} \left[e^{2.225/T}-1+ \frac{1}{\pi T} \int_0^\infty dE e^{-E/T} \delta_{p,n}(E)\right]
\end{equation}
(sum of all phase shifts in c.m. system). The reduction factor follows as 
\begin{equation}
R^{\rm vir}_d(T) =\frac{2^{1/2}}{3} b_{pn}(T) e^{-2.225/T}.
\end{equation}
Using the values given in \cite{HS} we find $R^{\rm vir}_d(1)=0.9883, \qquad R^{\rm vir}_d(2)=0.9453, \qquad R_d^{\rm vir}(3)=0.9004$, see also Tab. \ref{Tab:virpn}.
The effective binding energy follows as 
\begin{equation}
B^{\rm eff}_d(T) = -E_d^{\rm eff}(T) = -T \ln\left[\frac{2^{1/2}}{3} b_{pn}(T)\right].
\end{equation}

Using the $n - n$ scattering data, the values for $b_{n}(T)$ (full) are given in Ref. \cite{HS},
\begin{equation}
b_{n}(T)= \frac{2^{1/2}}{\pi T} \int_0^\infty dE e^{-E/T} \delta_{n,n}(E)-\frac{1}{2^{5/2}}.
\end{equation}
The effective binding energy follows as 
\begin{equation}
B^{\rm eff}_n(T) = -E_n^{\rm eff}(T) = -T \ln\left[\frac{1}{2^{1/2}}\left( b_{n}(T)+\frac{1}{2^{5/2}}\right)\right]
\end{equation}
and the reduction factor 
\begin{equation}
R^{\rm vir}_n(T) =\frac{1}{\pi T}  \int_0^\infty dE e^{-E/T} \delta_{n,n}(E).
\end{equation}
Values are shown in Tabs. \ref{Tab:virpn}, \ref{Tab:virnn}. The superscripts $a, r_0$ denotes scattering phase shifts taken from the scattering length and the effective range.
The tildes values are calculated with the quasiparticle correction, see \cite{R20}.

\begin{table}
\begin{center}
\hspace{0.5cm}
 \begin{tabular}{|c|c|c|c|c|c|c|c|}
\hline
$T$&$b_{pn}(T)$ \cite{HS} &$E_d^{\rm eff}(T)$ & $R_d^{\rm eff}(T)$&$E_d^{a,r_0}(T)$ & $R_d^{a,r_0}(T)$&$\tilde E_d^{a,r_0}(T)$ & $\tilde R_d^{a,r_0}(T)$\\
\hline
1& 19.4 & -2.2132 & 0.9883 	&-2.1953  &0.970737  & -2.2089 & 0.984028 \\
2 & 6.10 & -2.1125 & 0.94530 	&-1.98559  &0.887183  & -2.05406 &0.918079 \\
3 & 4.01 & -1.9103 & 0.9004 	&-1.61018   &0.814696  &-1.72622  &0.846825 \\
4 & 3.19 & -1.6319 & 0.8622 	&-1.11119   &0.756954  &-1.24862  &0.783414  \\
5 & 2.74 & -1.2796 & 0.8277  	&-0.508454   &0.709419  &-0.640388  &0.728387\\
6 & 2.46 & -0.8887 & 0.8003 	&0.189081    &0.668749  &0.0851844  &0.68043 \\
7 & 2.26 & -0.4433 & 0.7753  	&0.976248   &0.632977  &0.918021  &0.638264 \\
8 & 2.11 &  0.0428 & 0.7532  	&1.84889   &0.600954  &1.84983  &0.600883\\
9 &  2.00 & 0.5300 & 0.7363  	&2.80304   &0.571969  &2.87347  &0.56751 \\
10 & 1.91 & 1.0493 & 0.7208 	&3.83476    &0.545542  &3.98262  &0.537535 \\
\hline
 \end{tabular}
\caption{Effective bound state energy and virial correction expressed by the multiplying factor $R_d^{\rm eff}(T)$ in the $n - p$ channel containing the deuteron as bound state.}
\label{Tab:virpn}
\end{center}
\end{table}

\begin{table}
\begin{center}
\hspace{0.5cm}
 \begin{tabular}{|c|c|c|c|c|c|c|}
\hline
$T$&$b_{n}(T)$ \cite{HS} &$E_n^{\rm eff}(T)$ &$E_n^{a,r_0}(T)$ & $R_n^{a,r_0}(T)$&$\tilde E_n^{a,r_0}(T)$ & $\tilde R_n^{a,r_0}(T)$\\
\hline
1 & 0.288 &1.11277 & 1.05942 & 0.346657 & 1.5096 & 0.220999 \\
2 & 0.303 &2.16202 &2.12001&0.346454 & 3.0388& 0.218843  \\
3 & 0.306 &3.22432 & 3.23552 &0.340103 &4.69126 & 0.209349  \\
4 & 0.307 &4.29082 & 4.41082 &0.331972 &6.45687 & 0.199046  \\
5 & 0.308 &5.3532 & 5.65598 &0.322647 &8.33014 & 0.188996  \\
6 & 0.308 &6.3532 &6.97922  &0.312484 & 10.3071& 0.179451  \\
7 & 0.308 &7.3532 & 8.38453&0.30186 & 12.3835& 0.170492  \\
8 & 0.309 &8.54864 & 9.87243 &0.29111 &14.5542 & 0.162142  \\
9 & 0.310 &9.59871 & 11.4413 &0.280479 &16.8142 & 0.154394  \\
10&0.311 &10.6447 & 13.0883 & 0.270136 & 19.1583 & 0.14722  \\
\hline
 \end{tabular}
\caption{Effective bound state energy and virial correction expressed by the multiplying factor $R_n^{\rm eff}(T)$ in the $n - n$ channel .}
\label{Tab:virnn}
\end{center}
\end{table}

For $^5$He we consider
\begin{equation}
b_{\alpha n}(T)=\left(\frac{5}{4} \right)^{3/2} \frac{1}{\pi T} \int_0^\infty dE e^{-E/T} \delta_{\alpha n}(E)
\end{equation}
(c.m. system) also considered in \cite{HS}. 
The reduction factor is 
\begin{equation}
R_{^5{\rm He}}^{\rm vir}(T)= \left(\frac{4}{5} \right)^{3/2} \frac{1}{4} b_{\alpha n}(T)\,\,e^{28.3/T-27.56/T} 
\end{equation}
(the degeneracy factor 1/4 follows from the degeneracy factor in the phase shifts). This is the factor to multiply the NSE ground state contribution.
Using the values given in \cite{HS} we find $R_{^5{\rm He}}^{\rm vir}(1)=0.5661, \qquad R_{^5{\rm He}}^{\rm vir}(2)=0.5853, \qquad R_{^5{\rm He}}^{\rm vir}(3)=0.5883$. 
  
For $^8$Be we consider
\begin{equation}
b_{\alpha}(T)=2^{3/2} \frac{1}{\pi T} \int_0^\infty dE e^{-E/T} \delta_{\alpha}(E)+\frac{1}{2^{5/2}}
\end{equation}
(c.m. system) also considered in \cite{HS}. 
The reduction factor is 
\begin{equation}
R_{^8{\rm Be}}^{\rm vir}(T)=\left(\frac{1}{2} \right)^{3/2} \left[b_{\alpha}(T)-\frac{1}{2^{5/2}}\right]\,e^{56.6/T-56.496/T}.
\end{equation}
Using the values given in \cite{HS} we find $R_{^8{\rm Be}}^{\rm vir}(1)=0.9894, \qquad R_{^8{\rm Be}}^{\rm vir}(2)=1.46855,
 \qquad R_{^8{\rm Be}}^{\rm vir}(3)=1.9997$.
 
Calculations based on a separable potential approach have been performed in \cite{R20}.
Using the expressions given there, the reduction factor for $^4$H is 0.0654. Considering only the $P_{3/2}$ channel, 
 the reduction factor for $^5$He is 0.70716.

The calculation of the contribution of the continuum is more complex for the remaining isotopes.
Calculations with separable potentials are possible, but the potential parameters must be fitted to 
scattering data. We assume that the situation with $^7$He and $^9$He is comparable to $^5$He and take a similat 
reduction factor. Because these isotopes give only small contributions, a rough estimate is sufficient.

We have also to estimate the contribution of scattering states for isotopes with stable (with respect to strong  interaction)
ground states, as familiar from the virial expansion. This is known for $^2$H where we can use the result for the virial expansion $R^{\rm vir}_d(1.4)=0.9711$ given above.
For $^4$He, but also for $^3$H, $^3$He, the scattering state contributions are irrelevant because the continuum threshold ($S_n$) is high. 
For $^6$He, $^8$He, where we have also a considerable reduction  $R^{\gamma}_{A,Z}$, we expect  a significant contribution from continuum correlations. We consider
\begin{equation}
R^{\rm vir}_{A,Z}(T)= 1-e^{-S_n/T}\left[1-\frac{1}{\pi T}\int_0^\infty dE e^{-E/T}\delta_{A,Z}(E)\right]
\end{equation}
with scattering phase shifts $\delta_{A,Z}(E)$ as function of the energy $E$ in the c.m. system.
We can adapt phase shifts from other cases such as $n - p$ or $\alpha - n$ scattering, or perform calculations
with a separable potential (e.g. $^6$He, with $\gamma= 1.791$ fm$^{-1}$ as for $^5$He, but  $\lambda = 788.9$ instead of 670 MeV fm$^3$ to reproduce the bound state energy) to estimate this contribution. 
The phase shifts are only weakly decreasing in the low-energy region $\approx 1$ MeV of relevance, we obtain
$R_{^6{\rm He}}^{\rm vir}(1.2 \,{\rm MeV}) =0.9419,\,\,R_{^8{\rm He}}^{\rm vir}(1.2\, {\rm MeV}) =0.9841.$

\end{document}